# Negative Poisson's Ratio in Phagraphene

L.A. Openov and A.I. Podlivaev

National Research Nuclear University "MEPhI", Moscow 115409, Russian Federation

E-mail address: LAOpenov@mephi.ru

## ABSTRACT

We present the results of numerical simulation of elastic properties of phagraphene, a recently predicted but not synthesized yet quasi-two-dimensional allotrope of graphene. We show that the Poisson's ratio is positive for the planar configuration of phagraphene and negative for the nonplanar one. Both the Poisson's ratio and the Young's modulus are isotropic for the planar phagraphene and strongly anisotropic for the nonplanar phagraphene.

## 1. INTRODUCTION

The response of isotropic materials to mechanical stress is completely described by two scalar quantities, the Young's modulus $Y$, which characterizes the ability of a material to withstand longitudinal tension or compression, and the Poisson's ratio $\nu$, which is equal to the ratio of the transverse compressive strain to the longitudinal tensile strain [1]. The elastic properties of anisotropic solids are described using a more sophisticated mathematical apparatus: In general case, the equations of the linear theory of elasticity contain the tensor of elastic moduli. Nevertheless, often it is possible to restrict ourselves to just two parameters, $Y$ and $\nu$. In particular, for a quasi-two-dimensional graphene monolayer [2], the following values of the Young's modulus and the Poisson's ratio have been reported in the literature: $Y \sim 1.0$ TPa [3, 4] and $\nu = 0.15$-$0.45$ (see [5] and references therein).

For the majority of materials, the Poisson's ratio is positive. However, there are some exceptions, including fractal structures and other heterogeneous media (see [6] and references therein). Recently, it has been shown that for sufficiently narrow weakly deformed graphene nanoribbons the Poisson's ratio $\nu$ can be negative and reach the value of $\sim -1.5$ [7]. The reason for this effect is that compressive edge stresses lead to the formation of transverse displacement waves localized near the boundaries of the sample [8]. When the sample is stretched in the longitudinal direction, the wavy portions are straightened, and the sample is expanded (rather than compressed) in the transverse direction, which leads to a negative value of $\nu$.

As shown in Ref. [9], phagraphene – a recently predicted planar allotrope of graphene [10] – is unstable with respect to transverse atomic displacements, which generates the transverse displacement waves with an amplitude of $\sim 1$ Å and results in the formation of a nonplanar configuration of phagraphene [9, 11]. By analogy with graphene nanoribbons [7, 8], it can be expected that the Poisson's ratio for a nonplanar phagraphene will be negative. It should be noted, however, that there is a difference between graphene ribbons and phagraphene. While the negative Poisson's ratio for graphene ribbons is determined by boundary effects, in phagraphene it is caused by bulk effects (transverse displacement waves propagate throughout the entire sample).

The purpose of our work is to perform numerical calculations of the Poisson's ratio for planar and nonplanar configurations of phagraphene, as well as to calculate the Young's modulus for these configurations.

## 2. COMPUTATIONAL DETAILS

The planar phagraphene was simulated by a rectangular supercell consisting of 4 x 4 = 16 primitive 20-atom cells (Figs. 1a, 1b). In this supercell, the formation of transverse displacement waves led to a decrease in the energy of the supercell. The minimum energy was observed in the supercell with two waves (Figs. 2a, 2b), which we used to simulate the nonplanar configuration of phagraphene. We chose the periodic boundary conditions in both directions within the (*XY*) plane and the free boundary conditions in the transverse direction *Z*.

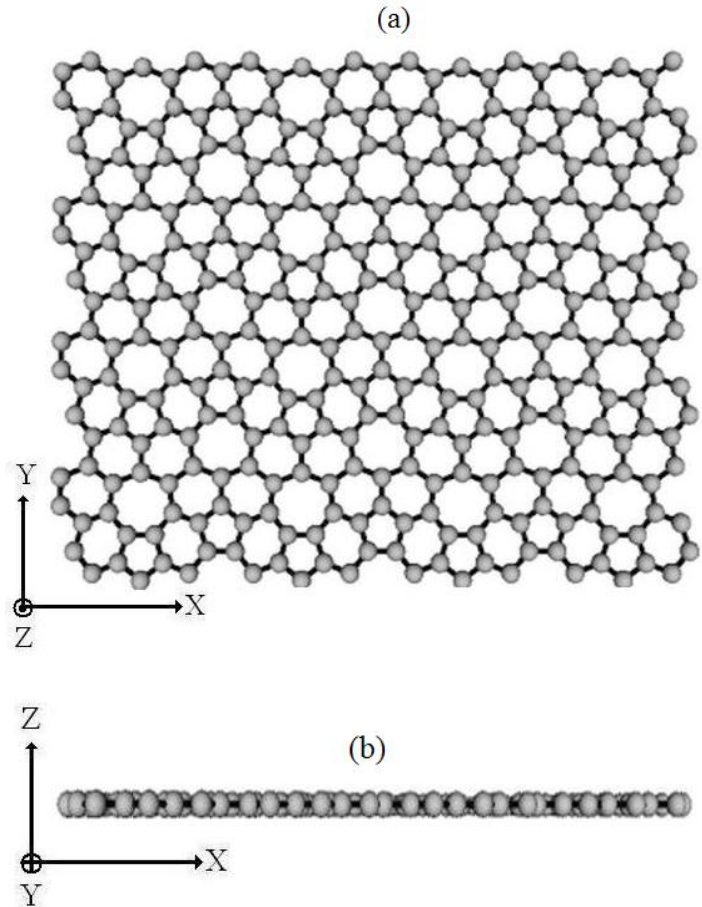

**Figure 1.** A 320-atom supercell for the planar configuration of phagraphene. Top view (a) and side view (b).

The energy of the supercell with specified atomic coordinates was calculated in the framework of the nonorthogonal tight-binding model [12], which takes into account all four valence orbitals of each carbon atom and has been shown to work well for the simulation of graphene and other carbon structures (see [9, 11, 13] and references therein).

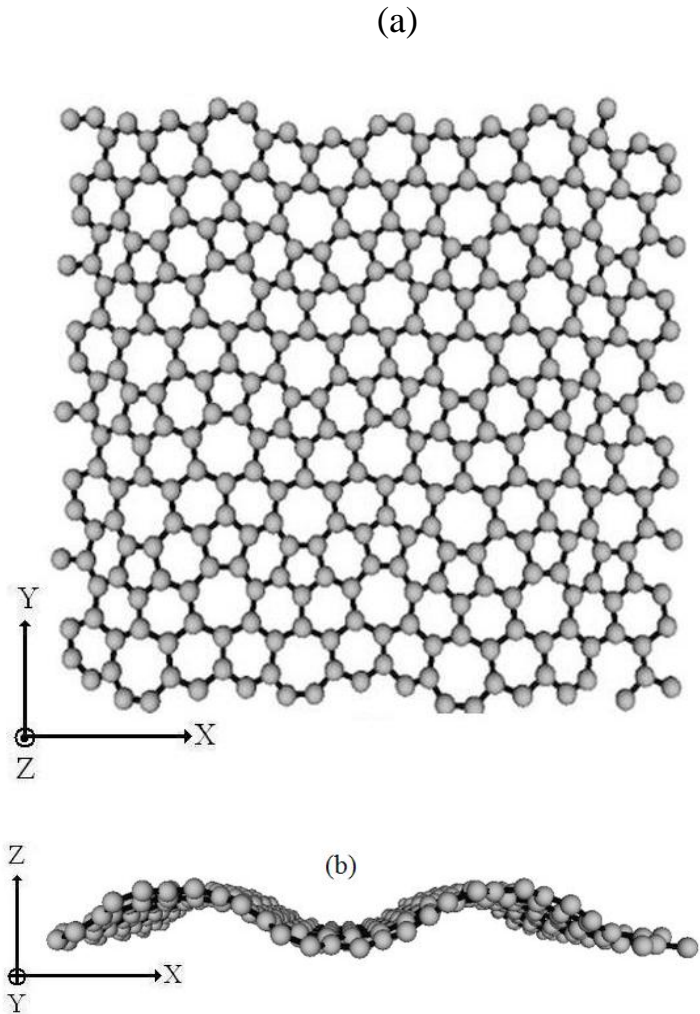

**Figure 2.** The same as in Figure 1, for the nonplanar configuration of phagraphene. The energy of this supercell is by 3.1 eV (~0.01 eV/atom) lower than that of the planar supercell.

# 3. RESULTS AND DISCUSSION

We start with formulas for the calculation of the Poisson's ratio ν and the Young's modulus $Y$ for quasi-two-dimensional materials which take into account the possible anisotropy of these quantities in the ($XY$) plane [14]. Let us assume that the sample has a rectangular shape with initial length $L_0$ (along the $X$ axis) and width $W_0$ (along the $Y$ axis). If, after stretching the sample along the $X$ axis, its length and width become equal to $L$ and $W$, respectively, the Poisson's ratio in the $Y$ direction can be calculated according to the formula

$$\nu_{YX} = -\varepsilon_Y / \varepsilon_X, \qquad (1)$$

where $\varepsilon_X = (L-L_0)/L_0$ and $\varepsilon_Y = (W-W_0)/W_0$ are the relative strains of the sample in the $X$ and $Y$ directions, respectively. Similarly, the Poisson's ratio in the $X$ direction after stretching the sample along the $Y$ axis is determined as

$$\nu_{XY} = -\varepsilon_X / \varepsilon_Y. \qquad (2)$$

In our case, the length and the width of the sample are equal to the periods of the supercell $a$ and $b$, respectively. It should be noted that, when calculating the Poisson's ratio $\nu_{YX}$, the strain $\varepsilon_X$ of the sample is specified as an input parameter, while the strain $\varepsilon_Y$ is found by minimization of the supercell energy, whereas in the calculation of $\nu_{XY}$, on the contrary, the strain $\varepsilon_Y$ is fixed, while the strain $\varepsilon_X$ is an output parameter and should be determined. Generally, the Poisson's ratio ν of the sample depends on the strain [7]. However, we restrict ourselves to the case of small strains and, for certainty, take the input parameters $\varepsilon_X$ and $\varepsilon_Y$ to be equal to 0.001 in all calculations.

Under weak deformations, the Young's moduli in the $X$ and $Y$ directions are calculated, respectively, according to the following formulas (similar to the case of nanotubes [15]):

$$Y_X = \frac{2\Delta E}{Sd\varepsilon_X^2}, \quad Y_Y = \frac{2\Delta E}{Sd\varepsilon_Y^2}, \qquad (3)$$

where $S = ab$ is the area of the supercell, $\Delta E$ is the increment of the energy of the supercell under the corresponding deformation, and $d$ is the thickness of the monolayer. Here it should be kept in mind that, the concept of the thickness of a sample does not have a clear meaning for quasi-two-dimensional systems, Nevertheless, this quantity has often been used to express the Young's modulus in conventional units (Pa). In particular, for graphene and carbon nanotubes, the sample thickness is usually taken to be 3.35 Å [3, 15], i.e., the distance between the adjacent graphene layers in graphite. We also set $d = 3.35$ Å in expressions (3). This makes it possible to compare the mechanical stiffnesses of phagraphene and graphene.

For the verification of our computational algorithms and the subsequent comparison with the available data on graphene, in the first stage we calculated the Poisson's ratio $\nu$ and the Young's modulus $Y$ in single-layer graphene making use of both rectangular and rhombic supercells. We found $\nu_{XY} = \nu_{YX} = 0.35$ and $Y_X = Y_Y = 0.97$ TPa, which agree well with the values $\nu = 0.34$ [7] and $Y = 1.0$ TPa [3] present in the literature.

Despite the apparent anisotropy of the planar configuration of phagraphene (Fig. 1a), we obtained for it (as for graphene) equal (within the limits of computational error) values of the Poisson's ratios $\nu_{XY} = \nu_{YX} = 0.38$ and very close values of the Young's moduli $Y_X = 0.84$ TPa and $Y_Y = 0.86$ TPa. Thus, elastic characteristics of the planar phagraphene differ little from graphene. The slightly less (by ~15%) value of the Young's modulus for the planar phagraphene, as compared to graphene, is apparently due to the fact that the planar phagraphene contains pentagons and heptagons formed by the C-C bonds, which are absent in graphene and which make the two-dimensional crystal lattice more "soft." For practical applications, this is not very important. Indeed, of primary interest are not elastic but electronic characteristics of planar phagraphene, which are determined by the presence of the so-called Dirac cones in its band structure [10].

The anisotropy of the atomic structure of the nonplanar phagraphene in the (*XY*) plane is associated with the presence of transverse displacement waves in it, the minima and maxima of which alternate along the *X* axis (Fig. 2b). This gives rise to a strong anisotropy of the Poisson's ratio ν and the Young's modulus *Y* after stretching the sample along the *X* and *Y* axes. Thus, we obtained the Poisson's ratios $\nu_{YX}$ = -0.04 and $\nu_{XY}$ = -0.50 and the Young's moduli $Y_X$ =0.06 TPa and $Y_Y$ = 0.75 TPa. Note that, despite the almost tenfold difference in the absolute values of $\nu_{XY}$ and $\nu_{YX}$, both these quantities are negative, whereas they are positive in the planar phagraphene. The reason for the abnormally low Young's modulus of the nonplanar phagraphene along the *X* axis lies in the corresponding orientation of the transverse displacement waves softening the sample in this direction (Fig. 2b). The Young's modulus of the nonplanar phagraphene along the *Y* axis is an order of magnitude larger, though slightly smaller, than that in graphene and planar phagraphene.

# 4. CONCLUSIONS

The calculations of the Poisson's ratio and the Young's modulus of phagraphene show that its elastic characteristics are isotropic in the planar configuration and strongly (tenfold) anisotropic in the nonplanar one. The main result is that the Poisson's ratios in the planar and nonplanar phagraphenes have different signs, regardless of the direction of the deformation. This can be used in practice to determine the type of atomic configuration of phagraphene samples.

*Acknowledgments.* This work was supported by the Russian Foundation for Basic Research (project 15-02-02764).